# Length, Weight, and Yield in Channel Catfish, Lake Diane, MI

Elizabeth Keenan, Sarah Warner, Ashley Crowe, and Michael Courtney
U.S. Air Force Academy, 2354 Fairchild Drive, USAF Academy, CO 80840
Michael.Courtney@usafa.edu

**Background:** Channel catfish (*Ictalurus punctatus*) are important to both commercial aquaculture and recreational fisheries. Little published data is available on length-weight relationships of channel catfish in Michigan. Though there is no record of public or private stocking, channel catfish appeared in Lake Diane between 1984 and 1995 and it has developed into an excellent fishery.

**Materials and Methods:** Sport angling provided 38 samples which were weighed and measured (fork length). Fillets were also weighed. The best fit estimates of parameters a and b in the model, $W(L) = aL^b$, were obtained by both linear least-squares (LLS) regression ($\log(W) = \log(a) + b \log(L)$) and non-linear least-squares (NLLS) regression. Best-fit parameters of an improved model, $W(L) = (L/L_1)^b$, were also determined by NLLS regression; the parameter $L_1$ is the typical length of a fish weighing 1 kg. The resulting best-fit parameters, parameter standard errors, and covariances are compared between the two models. The average relative weight for this sample of channel catfish is also determined, along with the typical meat yield obtained by filleting.

**Results:** NLLS regression yields parameter estimates of b = 3.2293 and a = 0.00522. The improved model yields the same estimate for the exponent, b, and a length estimate (parameter $L_1$) of 45.23 cm. Estimates of uncertainty and covariance are smaller for the improved model, but the correlation coefficient is r = 0.995 in both cases. LLS regression produced different parameter values, a = 0.01356 and b = 2.9726, and a smaller correlation coefficient, r = 0.980. On average, catfish in the sample weighed 106.0% of the standard weight, (Brown et al.) and the linear regression (no slope) of fillet yield vs. total weight suggests a typical fillet yield of 28.1% with r = 0.989.

**Conclusion:** Most of the fish in the sample were above the standard weight, heavier than the 75th percentile for their length. Channel catfish are doing well in Lake Diane and the population is well matched to the food supply. Management should attempt to maintain current population levels. In this case, the improved length-weight model, $W(L) = (L/L_1)^b$, provided lower uncertainties in parameter estimates and smaller covariance than the traditional model.

## I. Introduction

Lake Diane, Michigan, is on the southern edge of Hillsdale County. It is muddy due to clay particles suspended in the water. It has an average depth of 3 m and a maximum depth of 16 m. Lake Diane is a man-made reservoir of approximately 1.15 square km and was created in 1966 by damming a tributary to Clark Fork Creek, and has many permanent houses on the shoreline. Earlier studies have consistently shown stunted growth in black crappies (*Pomoxis nigromaculatus*) and bluegill (*Lepomis macrochirus*) attributed to an overabundance of these smaller species and a lack of piscivores. Between 1984 and 1991 an average of 2500 tiger muskellunge (*Esox masquinongy x lucius*) fingerlings were stocked eack year, but this program was unsuccessful in producing a successful fishery or bringing balance to the over-abundant (thus stunted) crappies and bluegill. A walleye stocking program recommended in 2003 is showing some signs of success. (Braunscheidel 2003)

Channel catfish were introduced to Lake Diane in an unknown manner some time between surveys in 1984 and 1995, when only 15 channel catfish were found. A 2001 survey found 393 channel catfish, 91% of which exceeded 30 cm in length. In the 2001 survey, channel catfish composed 36% of the weight of all species of fish sampled. Channel catfish were present in all age groups from 1 to 10 years, indicating a strong likelihood of natural reproduction, and even though they were growing slowly, the Michigan Department of Natural Resources concluded that Lake Diane is an excellent catfish fishery.

## II. Method

Catfish specimens were caught by means of sport angling using night crawlers or small fish as bait. The length of the fish was recorded as the distance from the front of the mouth to the fork of the tail with an accuracy of 0.2 cm. Each of 38 specimens were also weighed to an accuracy of 0.01 kg. After weighing and measuring, the fish were filleted and the weight of fillets was also recorded.

Length-weight relationships in fish traditionally employ the model, $W(L) = aL^b$, where L is length and W is weight. The parameters a and b were obtained by both linear least-squares (LLS) regression ($\log(W) = \log(a) + b \log(L)$) and non-linear least-squares (NLLS) regression. Best-fit parameters of an alternate model, $W(L) = (L/L_1)^b$, are also determined by NLLS regression, the parameter $L_1$ is the typical length of a fish weighing 1 kg. The resulting best-fit parameters, parameter standard errors, and covariances are compared between the two models. The average relative weight for this sample of channel catfish was also determined, (Brown et al. 1995) along with the typical meat yield percentage obtained by filleting.

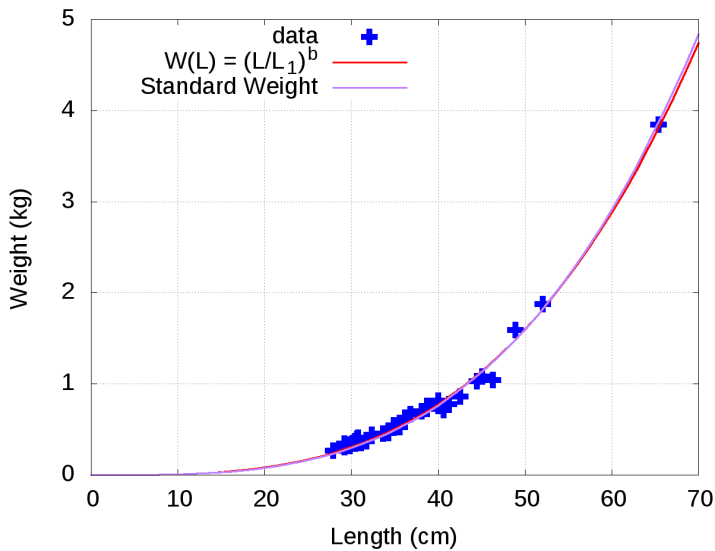

*Figure 1: (left) Channel catfish length vs. weight for fish with fork lengths of 28-65 cm. The best fit improved model is shown along with the standard weight curve. (Brown et al. 1995)*

| NLLS improved $W(L)=(L/L_1)^b$ | $L_1$ (cm) | 43.2344 |
|---|---|---|
| | $L_1$ error | 0.40% |
| | b | 3.2293 |
| | b error | 1.37% |
| | covariance | 0.5720 |
| | r | 0.9953 |
| NLLS traditional $W(L)=aL^b$ | a | 0.00522 |
| | a error | 17.29% |
| | b | 3.2293 |
| | b error | 1.35% |
| | covariance | -0.9980 |
| | r | 0.9953 |
| LLS traditional $W(L)=aL^b$ | a | 0.01356 |
| | a error | 4.81% |
| | b | 2.9726 |
| | b error | 2.19% |
| | covariance | -0.9990 |

*Table 1: Best-fit parameters and uncertainties for traditional and improved models.*

### III. Results

Figure 1 shows the data along with the best-fit for the improved model which is very close to the standard weight curve (Brown et al. 1995) so there is some overlap. Table 1 compares NLLS fit results to the traditional and improved models, along with the LLS result for the traditional model. For this data set, there are significant differences between LLS and NLLS results for the traditional model, and the improved model yields the same exponent as NLLS regression of the traditional model.

The improved model uses a parameter, $L_1$, which is the typical length of a fish weighing 1 kg. The improved model has about the same correlation coefficient and estimated uncertainty in the parameter b compared with the traditional model, but the uncertainty in $L_1$ is much smaller (0.40%) than the estimated uncertainties in the coefficient a using either the NLLS regression of linear least-squares (LLS) regression of log(W) vs. log(L). The covariance between parameters in the improved model is also smaller in magnitude than the covariance between parameters in the traditional model.

On average, catfish in the sample weighed 106.0% of the standard weight. (Brown 1995) Figure 2 shows a downward trend for relative weight vs. length for lengths from 25-45 cm, with only three large specimens above 45 cm showing relative weights near or above the standard weight. Linear regression (no slope) of fillet yield vs. total weight suggest a typical fillet yield of 28.1% with r = 0.989. Yield percentage does not depend strongly on weight (r = -0.192), length (r = -0.300), or relative weight (r = 0.190).

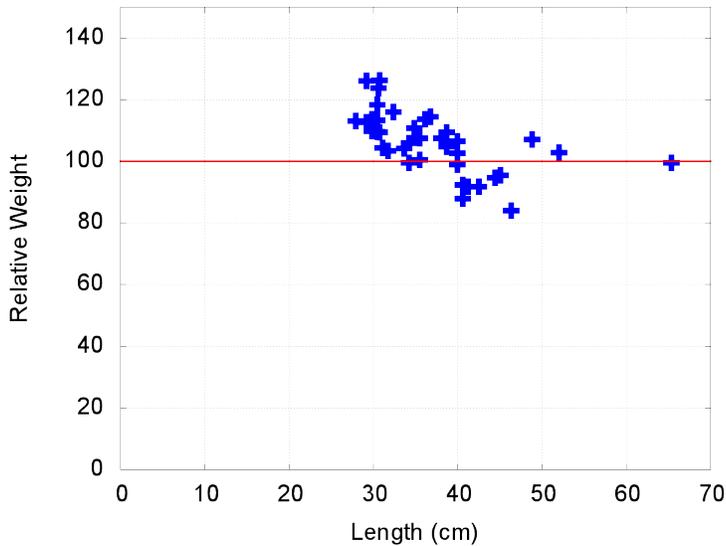
*Figure 2: Relative weight vs. length for channel catfish.*

**IV. Discussion**
Most of the fish in the sample were above the standard weight, meaning they were heavier than the 75th percentile for their length.  Many fish were close to the standard weight curve, and in Figure 1, the standard weight curve nearly overlaps the best-fit curve.  Channel catfish are doing well in Lake Diane and the population is well matched to the food supply, which probably includes a significant component of the stunted bluegill and crappies in catfish large enough to prey on these species.

Management should attempt to maintain current population levels or perhaps attempt to increase the population of channel catfish if needed to better balance the bluegill and crappies.  However, increasing the walleye population may be a better choice, as a more sight-based predator is somewhat complementary to the role of the channel catfish.  It may be that the standard weight of channel catfish tends to decrease between 25 cm and 45 cm because they cannot make good use of the stunted bluegill and crappies until over 45 cm in length.  When angling using small bluegill as bait, walleye in the 25-60 cm length range were much more commonly caught than channel catfish, so the walleye might be the more effective predator in lake conditions.

Channel catfish are important in aquaculture. (Wellborn 1988)  The fillet yield of 28% reported here is lower than the 40-45% suggested by Morris (1993).  The method of filleting was not specified in Morris (1993).  In the present study, the skin was first removed by making starter cuts in the skin behind the gills and along the back and pulling the skin off with pliers, as is the custom in the southern United States.  A fillet knife is then used to cut off the fillets working as closely as possible to the spine and ribs, but excluding all bones on the first cut.

For channel catfish in our sample, the improved length-weight model, $W(L) = (L/L_1)^b$, provided lower uncertainties in parameter estimates and smaller covariance than the traditional model.  The estimate of parameter, $L_1$ = 43.23 cm, as the length of a fish weighing 1 kg makes sense, as most catfish weighing close to 1 kg are close to this length.  Thus the parameters from the improved model are more intuitive than the coefficient a in the traditional model.

## V. Acknowledgements
The authors acknowledge helpful discussions with Amy Courtney, PhD (BTG Research) and Beth Schaubroeck, PhD (USAFA Department of Mathematical Sciences). We also thank BTG Research for supporting this work and E.R. Courtney (BTG Research), a valuable research assistant, for help weighing and measuring.